\documentclass[twocolumn,twoside,slac_two]{revtex4}
\usepackage{graphicx}
\usepackage{fancyhdr}
\begin{document}

\title{Reach of Future Non-accelerator Neutrino Efforts}

%

\author{R. Henning}
\affiliation{University of North Carolina at Chapel Hill and Triangle Universities Nuclear Laboratory}

\begin{abstract}
We present a brief review of the current understanding of neutrino flavor mixing and masses, followed by a discussion of the current and future experimental programs in neutrinoless double-beta decay, direct neutrino mass measurements, indirect neutrino mass determination from cosmology, solar neutrinos and other probes.  We emphasize how these programs will improve our understanding of neutrino masses and flavor mixing.  
\end{abstract}

\maketitle

\thispagestyle{fancy}


\section{Introduction}
\label{se:intro}

There now exists compelling experimental evidence that propagating neutrinos undergo flavor oscillations. This can  be ascribed to the difference between the mass and flavor eigenstates of neutrinos. If we assume the existence of three neutrino flavors, then the relationship between flavor and mass eigenstates is given by the so-called neutrino or Pontecorvo-Maki-Nakagawa-Sakata (PMNS) mixing matrix that is similar to the CKM matrix of the quark sector. Like quarks, the existence of more than two flavors of neutrinos allow $CP$-violation in neutrino interactions, quantified by adding a complex phase to the PMNS matrix. $CP$-violation in neutrino interactions has not been observed, but could have significant implications in particle physics and  cosmology if it is finite. The PMNS matrix can be parameterized as:
\begin{equation}
\label{eq:PMNS_matrix}
 \begin{array}{l}
 U_{PMNS}= \left( \begin{array}{ccc}
1 & 0 & 0 \\
 0 & c_{23} & s_{23} \\
 0 & -s_{23} & c_{23} \\
 \end{array} \right)
 \left( \begin{array}{ccc}
c_{13} & 0 & e^{i \delta}s_{13} \\
 0 & 1 & 0\\
 - e^{i\delta} s_{13}& 0 & c_{13} \\
 \end{array} \right)
 \times \\
 \left( \begin{array}{ccc}
c_{12} & s_{12} & 0 \\
 -s_{12} & c_{12} & 0 \\
 0 & 0 & 1 \\
 \end{array} \right)
 \left( \begin{array}{ccc}
 e^{i\alpha_1} & 0 & 0 \\
 0 & e^{i\alpha_2} & 0 \\
 0 & 0 & 1 \\
 \end{array} \right)
\end{array}
\end{equation}
 where $s_{ij} = \sin{\theta_{ij}}$ and $c_{ij} = \cos{\theta_{ij}}$ are the mixing angles and $\delta$ is the $CP$-violating phase. If neutrinos are Majorana (see \S \ref{se:DBD}) then two additional $CP$-violating phases, the so-called Majorana phases $\alpha_1$ and $\alpha_2$, are allowed. The current experimental values of these parameters are given in Table~\ref{tab:current_knowledge}. All these measurements are from oscillation experiments that were insensitive to $CP$-violating effects. The mass ($\nu_1, \nu_2, \nu_3$) and flavor ($\nu_e, \nu_\mu, \nu_\tau$) eigenstates are now related by:
  
  \begin{equation}
  \label{eq:mass_flav_rel}
  \left( \begin{array}{c}
  \nu_e \\
  \nu_\mu \\
  \nu_\tau \\
  \end{array} \right) =U_{PMNS}
  \left( \begin{array}{c}
  \nu_1 \\
  \nu_2 \\
  \nu_3 \\
  \end{array} \right)
  \end{equation}

The rates of oscillations determine the absolute mass-squared differences between the mass eigenstates, typically reported as $\Delta m_{ij}^2 = m_i^2 - m_j^2$ , where $m_i$ is the mass of the $i$th neutrino mass eigenstate. Current values are reported in Table~\ref{tab:current_knowledge}. The absolute mass scale of neutrinos is much less certain and three different mass hierarchies are possible. They are quasi-degenerate ($m_1\approx m_2 \approx m_3$), normal ($m_1 \approx m_2 << m_3$), or inverted ($m_3 << m_1 \approx m_2$). It is the task of future experiments to determine the mass hierarchy of the neutrinos. 
 
 \begin{table}[h]
 \begin{center}
 \caption{Current knowledge of the neutrino mixing parameters of the PMNS matrix from~\cite{PDG06}. Also shown are the methods or experiments used to measure these quantities.}
 \begin{tabular}{|l|l|l|}
 \hline 
 \textbf{Parameter} & \textbf{Value} & \textbf{Experiment or} \\
 & & \textbf{Method} \\
 \hline
 $\mathrm{sin}^2(2\theta_{12})$ & $0.86^{+0.03}_{-0.04}$ & Solar and KamLAND \\
 \hline
  $\mathrm{sin}^2(2\theta_{23})$ & $ > 0.92$ & Atmospheric \\
 \hline
  $\mathrm{sin}^2(2\theta_{13})$ & $ < 0.19$ & Reactor (CHOOZ) \\
  \hline 
  $|\Delta m^2_{32}|$ & $1.9-3.0\times10^{-3}\,\mathrm{eV}^2$ & Super-K and MINOS \\
  \hline 
  $|\Delta m^2_{21}|$ & $8.0\pm0.3\times10^{-5}\,\mathrm{eV}^2$ & Solar and KamLAND \\
  \hline
  $\delta$ & unknown & Possibly future\\
  & & long-baseline\\
  \hline
  $\alpha_1, \alpha_2$ & unknown & Possibly \\
  & & double-beta decay\\
 \hline 
 \end{tabular}
 \label{tab:current_knowledge}
 \end{center}
 \end{table}

Some of the experimental methods listed in Table~\ref{tab:current_knowledge} require non-accelerator based techniques. These techniques and the range of future efforts are the subject of this paper and are described in subsequent sections. 

\section{Neutrinoless Double-beta Decay and Neutrino Mass}
\label{se:DBD}

Because neutrinos are electrically neutral, the only quantum number that can distinguish between neutrinos ($\nu$) and anti-neutrinos ($\bar{\nu}$) is lepton flavor number. However, there is no fundamental reason this quantity should be conserved, even in the standard model, and there are many extensions to the standard model that require that it be violated. In this case the distinction between $\nu$ and $\bar{\nu}$ is unclear and it becomes possible that the $\nu$ can be its own anti-particle  or a so-called Majorana fermion. Majorana fermions were first postulated by Ettore Majorana as solutions to the Dirac equation in 1937~\cite{majo37}. They are different from the more familiar Dirac fermions solutions that have distinct anti-particle states. Surprisingly, the experimental data is consistent with both Majorana and Dirac neutrinos and the determination of the nature of the neutrino is difficult due to the small neutrino masses and the handedness of the weak interaction. The observation of neutrinoless double-beta decay in an atomic nucleus ($0\nu\beta\beta$-decay) is currently the 
only practical way to show that the neutrino is Majorana.

$0\nu\beta\beta$-decay is a currently unobserved\footnote{There exists a highly controversial claim of a discovery. See~\cite{avig07} and references therein for a discussion of this claim.} nuclear decay where two neutrons in an atomic nucleus convert into two protons and electrons with no neutrinos emitted. The existence of this process implies the existence of a Majorana mass term in the neutrino Lagrangian that mixes neutrinos and anti-neutrinos, hence that neutrinos are Majorana fermions~\cite{sche82}. It is also obviously a $\Delta L=2$  lepton number violating process. A related process is two neutrino double-beta decay where two antineutrinos are emitted as well. This process has been observed in several nuclei, but it is an allowed second order weak process and does not imply that the neutrino is Majorana. 
Similar nuclear decays with no neutrino emission processes, such as double electron capture, double positron emission, or simultaneous electron capture and positron emission would also imply that the neutrino is Majorana. There are experimental efforts underway to search for these processes, but we will focus on $0\nu\beta\beta$-decay in this paper. Many nuclei can undergo $0\nu\beta\beta$-decay, but experimentalists prefer even-even nuclei that are also stable against normal beta decay, since the beta decay overwhelms the extremely slow $0\nu\beta\beta$-decay rate. 

Many processes beyond the standard model can mediate $0\nu\beta\beta$-decay, such as supersymmetry and right-handed currents~\cite{avig07}. It is likely that the dominant process is the exchange of a massive Majorana neutrino between the two neutrons. In this case the measured half-life of the decay provides a measurement of the absolute neutrino mass scale, as opposed to the neutrino mass squared differences from oscillation experiments. Specifically, 
\begin{equation}
\label{eq:mass_rel_DBD}
[T^{0\nu}_{1/2}]^{-1} = G^{0\nu}(E_0,Z)| \langle m_{\beta\beta} \rangle |^2|M^{0\nu}|^2
\end{equation}
where $[T^{0\nu}_{1/2}]^{-1}$ is the inverse of the measured half-life, $G^{0\nu}(E_0,Z)|$ is an exactly calculable phase space factor, $|M^{0\nu}|$ is the matrix element that describes the nuclear physics and $|\langle m_{\beta\beta}\rangle |$ the so-called effective Majorana electron neutrino mass. The latter term can be expressed in terms of neutrino mixing matrix elements and neutrino masses as:
\begin{equation}
\label{eq:mbb_expr}
|\langle m_{\beta\beta} \rangle| = | \sum_i |U_{ei}|^2m_{\nu_i}e^{i\alpha_i} |
\end{equation} 
It is clear that the measured half-life for $0\nu\beta\beta$-decay probes the absolute mass-scale of the neutrino. However, the nuclear matrix elements are difficult to compute and prone to large uncertainties. 

Experiments that search for $0\nu\beta\beta$-decay face many challenges. They require a significant reduction in ionizing radiation backgrounds, necessitating deep underground sites, special materials selection and handling, mitigation of the high cost of enriching isotopes and advanced analysis techniques.  The current generation of experiments use tens of kilograms of isotope, enriched in most cases. They will probe half-lives in the $10^{26}$ to $10^{27}$ year range and neutrino mass scales down to $100\,\mathrm{meV}$. Their costs are in the ten to twenty million dollars range. The next generation of experiments should be operational in about 10 years and scale the previous quantities by an order of magnitude. 
Many different experimental approaches are currently used to search for $0\nu\beta\beta$-decay.  Table~\ref{tab:exp_approaches} summarizes these techniques and lists current experiments underway or under development
 \begin{table}[h]
 \label{tab:exp_approaches}
 \begin{center}
 \caption{Selected current and future experiments that search for $0\nu\beta\beta$-decay. Also given are the experimental techniques and isotopes used.}
 \begin{tabular}{|l|l|}
 \hline
 \textbf{Experimental} &  \textbf{Experiments and} \\
  \textbf{Technique} & \textbf{Isotopes Used} \\
 \hline
 Cryogenic Bolometry & CUORE/Cuoricino ($^{130}$Te) \\
 \hline
 Scintillation & CAMEO ($^{116}$Cd), \\
 & CANDLES ($^{48}$Ca), EXO ($^{136}$Xe), \\
 &  SNO+ ($^{150}$Nd), XMASS ($^{136}$Xe) \\
\hline
 Ionization & COBRA (CdTe), \\
  & GERDA ($^{76}$Ge), \\
  &  {\textsc MAJORANA} ($^{76}$Ge) \\
  \hline
  Time Projection & MOON ($^100$Mo),  Nemo (many),  \\
  and Tracking &HPGeTPC ($^{136}$Xe)\\
 \hline
 \end{tabular}
 \end{center}
 \end{table}
 
\section{Direct Neutrino Mass Measurements}
\label{se:direct_mass_measurements}
The search of endpoint effects in the spectrum of electrons emitted during nuclear beta decay is a well-known technique to determine the mass of the neutrino~\cite{otte06}. Specifically, this technique measures the so-called effective electron neutrino mass, $\langle m_\beta \rangle$:
\begin{equation}
\label{eq:effective_nue_mass} 
\langle m_\beta \rangle = \sum_i |U_{ei}|^2m_i^2
\end{equation} 
Current state-of-the art techniques employ large magnetic spectrometers to measure the endpoint of tritium beta decay that has a favorably low $Q$-value of  $18.6\,\mathrm{keV}$. The current limit from this technique is~\cite{PDG06}:
\begin{equation}
\label{eq:tbdk_limit}
\langle m_\beta \rangle < 2.0\,\mathrm{eV}/c^2
\end{equation}
and is based on experiments performed at Troitsk~\cite{loba99} and Mainz~\cite{pica92}. 
A next generation spectrometer, the Karlsruhe Tritium Neutrino Experiment (KATRIN), is nearing completion and anticipates a five year run starting in 2009. KATRIN hopes to measure $\langle m_\beta \rangle$ or improve the current limit by a factor of ten to $0.2\,\mathrm{eV/c^2}$~\cite{katr04}.

\section{Indirect Limits from Cosmology}
\label{se:indirect_limits}
One of the predictions of the Big-Bang model is that the universe is permeated with cosmological relic neutrinos. These neutrinos have a thermal energy distribution at a temperature of $1.7\,\mathrm{K}$ and a number density of $\sim 300\,\mathrm{cm^{-3}}$. Free-streaming of these neutrinos suppresses structure formation in the universe, leading to a variety of observable effects that can be used to constrain the sum of the masses of the neutrinos. The current best constraint is from~\cite{selj06}. These authors combine results from observations of large scale structure, the Lyman-$\alpha$ forest, Supernovae Type Ia and recent cosmic-microwave background data. Assuming three neutrino states, their limit is:
\begin{equation}
\label{eq:cosmo_nu_mass}
m_1+m_2+m_3 < 0.17\,\mathrm{eV}
\end{equation}
This limit is quite constraining but is dependent on the cosmological model used by the authors. Direct measurement of the neutrino masses is still crucial to remove this uncertainty. They claim that they can improve their limit by a factor of two with improved understanding of systematics.  

\section{Combined Mass Limits}
\begin{figure}[h]
\centering
\includegraphics[width=80mm]{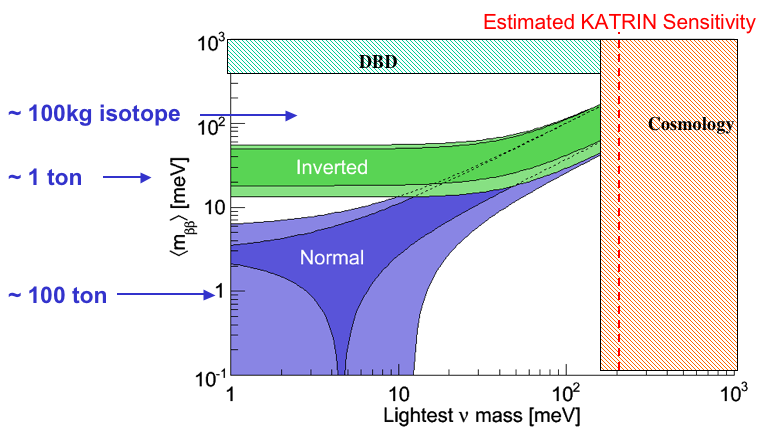}
\caption{Plot showing neutrino mass reaches of current and future experiments. The dark regions correspond to allowed regions given the current uncertainties in the mixing angle measurements. The lighter regions include arbitrary Majorana phases as well. Filled rectangular areas have been excluded by $0\nu\beta\beta$-decay experiments (DBD) and cosmological observations. The massed on the left are estimates of the amount of isotope required by a $0\nu\beta\beta$-decay experiment to probe a specific mass region. Plot courtesy of A.~G. Schubert (U. of Washington) and J. Detwiler (Lawrence Berkeley National Laboratory).} 
\label{fig:mass_reaches}
\end{figure}

The techniques discussed so far probe neutrino masses in different and complementary ways. Thes can be compared by plotting the currently excluded regions of lightest neutrino mass vs. effective Majorana neutrino mass, as shown in Figure~\ref{fig:mass_reaches}. 

\section{Solar Neutrinos}
\label{se:solar_nus}
\begin{figure}[h]
\centering
\includegraphics[width=80mm]{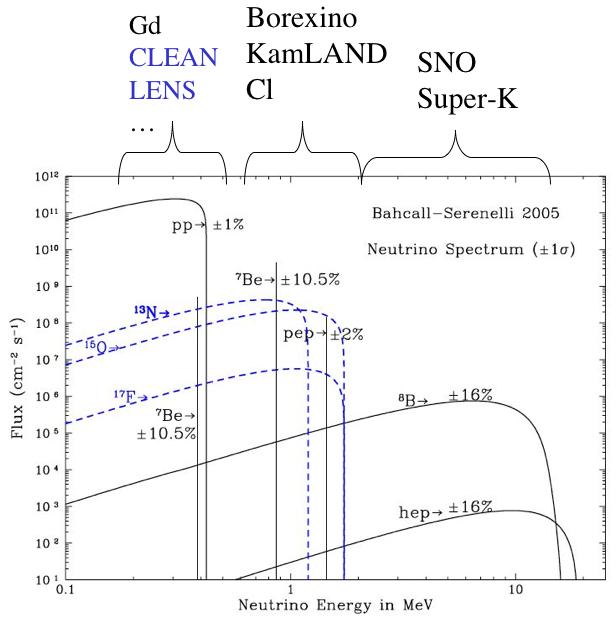}
\caption{The predicted solar neutrino flux from the sun, taken from~\cite{bahc05}. The neutrino fluxes from continuum sources are 
given in units of number $\mathrm{cm^{-2} s^{-1} MeV^{-1}}$ at one 
astronomical unit, and the line fluxes are given 
in number $\mathrm{cm^{-2} s^{-1}}$. The authors of this proceedings have also indicated the ranges covered by existing or previous experiments. The ''Gd'' and ''Cl'' refers to radiochemistry experiments.}
\label{fig:solar_nu}
\end{figure}

Solar neutrinos provided the first hint of neutrino oscillations with the famous Davis experiment in the Homestake mine in South Dakota. Since then several experiments, culminating with the Sudbury Neutrino Observatory (SNO), have confirmed the hypothesis that the paucity of electron neutrinos from the sun in comparison to solar model calculations is due to neutrino oscillations. These experiments also constrain the $\theta_{12}$ mixing angle by measuring the transition $\nu_e \rightarrow \nu_x$. The predicted energy spectrum of neutrinos produced in the sun are shown in figure~\ref{fig:solar_nu}. 
Current measurements of $\theta_{12}$ rely on the $^8\mathrm{B}$ flux intensity on earth that is described by the so-called Large Mixing Angle with matter effects. The matter effects are referred to as the Mikheyev-Smirnov-Wolfenstein (MSW) effect~\cite{wolf78,mikh85} and is caused by changes in the effective masses of neutrinos due to charged current interactions of the electron neutrinos with electrons in matter. The MSW-effect becomes negligible at lower energies and it is important to verify the solar models and improve our current measurements of $\theta_{12}$ by measuring the solar neutrino flux at lower energies. Two experiments, KamLAND-solar and Borexino are currently operational and measuring the $^7\mathrm{Be}$ flux. Proposed experiments on the 10 year timescale such as LENS and CLEAN will measure the pp solar neutrino flux in real time. This is particularly important for solar physics as well, since the pp process is the dominant source of neutrinos from the sun.

\section{Other Probes}
\label{se:other_probes}

Cosmic-ray neutrinos were important in verifying neutrino oscillations and also provide the current best limit on $\theta_{23}$~\cite{ashi05}. The next level of precision in determining $\theta_{23}$ will come from proposed long-baseline experiments that will also probe $CP$-violation in the neutrino sector. Ultra-high energy cosmic-ray have some capabilities to perform flavor physics, but these are not discussed here, as they are primarily of astrophysical interest. Coherent neutrino nuclear scattering has not been observed, but is a process that is well-understood in the Standard Model. The recent development of low-threshold detector technologies makes the measurement of this process achievable. Of course, any deviation observed from the Standard Model prediction of this cross-section would be indicative of new physics. 
 
\begin{acknowledgments}
The author wishes to acknowledge the support of the State of North Carolina, the US NSF (Award \# PHY-0705014) and the US DOE (Grant \# DE-FG02-97ER41041).
\end{acknowledgments}

\bigskip 

\bibliography{main}

\newcommand{\noopsort}[1]{} \newcommand{\printfirst}[2]{#1}
  \newcommand{\singleletter}[1]{#1} \newcommand{\switchargs}[2]{#2#1}
\begin{thebibliography}{10}

\bibitem{PDG06}
W.-M.~Yao et~al (Particle Data~Group).
\newblock {\em J. Phys. G: Nucl. Part. Phys.}, \textbf{~33~}:~1, 2006.

\bibitem{majo37}
E.~Majorana.
\newblock {\em Nuovo Cimento}, \textbf{~14~}:~171, 1937.

\bibitem{avig07}
F.~T.~Avignone et~al.
\newblock {\em Rev. Mod. Phy.}, \textbf{~80~}:~481, 2007.

\bibitem{sche82}
Schecter et~al.
\newblock {\em Phys. Rev. D}, \textbf{~25~}:~2951, 1982.

\bibitem{otte06}
J.~Bonn E.W.~Otten and Ch. Weinheimer.
\newblock {\em International Journal of Mass Spectrometry},
  \textbf{~251~}:~173, 2006.

\bibitem{loba99}
V.~M.~Lobashev et~al.
\newblock {\em Phys. Lett. B}, \textbf{~460~}:~227, 1999.

\bibitem{pica92}
A.~Picard et~al.
\newblock {\em Nucl. Instr. and Meth. B}, \textbf{~63~}:~345, 1992.

\bibitem{katr04}
KATRIN Collaboration.
\newblock Katrin design report 2004.
\newblock http://www-ik.fzk.de/tritium/publications/documents/
  DesignReport2004-12Jan2005.pdf.

\bibitem{selj06}
U.~Seljak et~al.
\newblock {\em JCAP}, \textbf{~0610~}:~014, 2006.

\bibitem{bahc05}
A.~M.~Serenelli J.~N.~Bahcall and S.~Basu.
\newblock {\em Astrophys. J.}, \textbf{~621~}:~L85, 2005.

\bibitem{wolf78}
L.~Wolfenstein.
\newblock {\em Phys. Rev. D}, \textbf{~17~}:~2369, 1978.

\bibitem{mikh85}
S.~P. Mikheev and A.~Yu. Smirnov.
\newblock {\em Sov. J. Nucl. Phys.}, \textbf{~42~}:~913, 1985.

\bibitem{ashi05}
Y.~Ashie et~al.
\newblock {\em Phys. Rev. D}, \textbf{~71~}:~112005, 2005.

\end{thebibliography}
\bibliographystyle{unsrt}


\end{document}